\documentclass[aps,prl,groupedaddress,fleqn,twocolumn]{revtex4}
\pdfoutput=1
\usepackage{graphicx}
\usepackage{amsmath}
\usepackage{gensymb}
\begin{document}

\title{Dynamical quantum indistinguishability}
\author{K. Trachenko$^{1}$}
\address{$^1$ School of Physics and Astronomy, Queen Mary University of London, Mile End Road, London, E1 4NS, UK}

\begin{abstract}
We observe that quantum indistinguishability is a dynamical effect dependent on measurement duration. We formulate a quantitative criterion for observing indistinguishability in quantum fluids and its implications including quantum statistics and propose a viscoelastic function describing both long-time and short-time regimes where indistinguishability and its implications are operative and inactive, respectively. On the basis of this discussion, we propose an experiment to observe a transition to the short-time non-equilibrium state where the implications of indistinguishability become inoperative.
\end{abstract}


\maketitle

Indistinguishability of identical particles in quantum mechanics has several implications including limiting the wavefunction to be symmetric or antisymmetric (related to Bose-Einstein or Fermi-Dirac statistics), Pauli principle and exchange interaction \cite{landau}. The wave function of two particles exchanging places must obey

\begin{equation}
\psi(\xi_1,\xi_2)=\pm\psi(\xi_2,\xi_1)
\label{psi}
\end{equation}

\noindent where $\xi$ are spatial and spin coordinates. A related effect is the emergence of the exchange integral $J_0$ in the average value of the interaction operator $U(r_2-r_1)$ between two electrons:

\begin{eqnarray}
\begin{split}
& \overline{U}=B\pm J_0\\
& B=\int U(r_2-r_1)|\phi_1({\bf r}_1)|^2|\phi_2({\bf r}_2)|^2dV_1dV_2\\
& J_0=\int U(r_2-r_1)\phi_1({\bf r}_1)\phi_1^*({\bf r}_2)\phi_2({\bf r}_2)\phi^*_2({\bf r}_1)dV_1dV_2\\
\end{split}
\label{exchange}
\end{eqnarray}

\noindent where $\phi_1$ and $\phi_2$ are wave functions of non-interacting particles, plus and minus signs correspond to symmetric and antisymmetric wave function with the total spin 0 and 1, respectively, $B$ is the additive constant and $\pm J_0$ is the energy level shift due to exchange \cite{landau}.


The textbook discussions of examples such as (1), (2) as well as many others where quantum statistics is important, tacitly assume that particles are able to physically exchange places. As we see below, this assumption may or may not be correct depending on an observable and measurement duration. For example, quantum statistics plays an decisive role in liquid 4He and 3He where atoms flow and are able to exchange places. However, thermodynamic properties in solid 3He and 4He are largely governed by lattice effects as in conventional solids \cite{dobbs}, with atomic exchanges representing a weak effect in comparison \cite{l2}. Therefore, the current discussion of quantum indistinguishability and its implications such as quantum statistics is incomplete. This was noted by Leggett who says that in order to observe the effects of indistinguishability and quantum statistics, particles need to physically exchange places and ``find out'' that they are indistinguishable, with consequences for the symmetrisation of the wave function \cite{l1,l2}. Leggett uses another example, the difference of vibration and rotation spectra of heteronuclear and homonuclear molecules, underlying a difference between rotation where identical particles physically change places and vibration where they do not \cite{l1,l2}.

Similarly to theoretical discussion, computational techniques tacitly assume particle exchange. Physical exchange of particles in embedded in path integral Monte Carlo simulations of liquid He where trajectories are sampled from the space of particle permutations consistent with Bose symmetry. Critical effects such as the superfluid transition are associated with a large number of particle exchanges visualised as winding paths of cyclic exchanges \cite{ceperley}. The exchange energy related to fermions plays an important role in electronic structure calculations, and is rigidly added to the total system energy \cite{payne}.

We observe that physical exchange of particles is a dynamical process with a characteristic time scale. Consequently, particle indistinguishability and its consequences depend on details of particle dynamics involved in setting an observable and measurement duration. To the best of our knowledge, this dependence has not been previously discussed. In this paper, we take a first step to provide a quantitative description of this dependence. We formulate a quantitative criterion for observing indistinguishability in quantum fluids and its implications including quantum statistics and propose a viscoelastic function describing both long-time and short-time regimes where indistinguishability and its implications are operative and inactive, respectively. We finally propose an experiment to observe a transition to the short-time non-equilibrium state where the implications of indistinguishability become inoperative.

We start a discussion with fluids where an important time scale is liquid relaxation time $\tau$, the time between two consecutive large flow-enabling particle displacements, jumps, from one quasi-equilibrium position to the adjacent position \cite{frenkel}. $\tau$ is governed by the activation energy barrier $\tilde{U}$ and varies in a very wide range: from typically 0.1 ps in high-temperature liquids to hours at the liquid-glass transition, i.e. by about 16 orders of magnitude. In between the jumps and at time $t\ll\tau$, liquid particles vibrate as in solids and are unable to exchange places. Therefore, a necessary condition for liquid particles to exchange places is that the time of an experiment where an observable is measured should exceed $\tau$:

\begin{equation}
t\gg\tau
\label{c1}
\end{equation}

In this case, a measured quantity would show the effect of particle indistinguishability and its implications such as Eqs. \eqref{psi}, \eqref{exchange} and other effects of quantum statistics.

Another way of writing this condition is

\begin{equation}
\omega\tau\ll 1
\label{cond}
\end{equation}

\noindent where $\omega$ is the frequency at which a long-time observable is measured.

Eq. \eqref{cond} is a familiar condition of hydrodynamic equilibrium \cite{frenkel}. The novelty here is its relation to the regime where particle indistinguishability and quantum statistics play a role.

A regime opposite to \eqref{c1} and \eqref{cond} is $t\ll\tau$, or $\omega\tau\gg 1$, where $\omega$ is the frequency at which an observable is measured during short times $t\ll\tau$. In this regime, particles do not exchange places, and implications of quantum indistinguishability are not operative. $\omega\tau\gg 1$ corresponds to a non-hydrodynamic solid-like regime of liquid dynamics \cite{frenkel}. This regime has important implications for liquid collective modes and thermodynamics \cite{ropp}.


It is interesting to ask what theory describes both long-time and short-time regimes of fluids dynamics as well as a transition between them and, consequently, indistinguishability and its measurable implications. A simple approach describing these effects is the Maxwell-Frenkel theory of liquids (see, e.g. Ref. \cite{ropp} for review). Maxwell proposed that liquid response to, for example, shear stress $P$ is neither hydrodynamic nor elastic, but has contributions from both effects \cite{maxwell}. Frenkel has made this more specific and wrote \cite{frenkel}:

\begin{equation}
\frac{dv}{dy}=\frac{ds}{dt}=\frac{P}{\eta}+\frac{1}{G}\frac{dP}{dt}
\label{a1}
\end{equation}

\noindent where $v$ is the velocity perpendicular to $y$-direction, $s$ is shear strain, $\eta$ is viscosity and $G$ is shear modulus.

According to Eq. (\ref{a1}), shear deformation in a liquid is the sum of the viscous and elastic deformations, given by the first and second right-hand side terms. This has given rise to term ``viscoelasticity'' of liquids. Both deformations are treated in (\ref{a1}) on equal footing, hence ``elastoviscosity'' would be an equally legitimate term.

Eq. \eqref{a1} gives a viscoelastic crossover: consider an external force $P\propto e^{i\omega t}$. Then, $\frac{dv}{dy}=\frac{1}{\eta}(1+i\omega\tau)P$ where $\tau=\frac{\eta}{G}$. $\omega\tau\ll 1$ gives viscous response, $\frac{dv}{dy}=\frac{P}{\eta}$, whereas $\omega\tau\gg 1$ gives elastic response: $\frac{ds}{dt}=i\omega GP=G\frac{dP}{dt}$. Augmenting this result with a microscopic theory, Frenkel identified liquid relaxation time introduced earlier with  $\tau\approx\frac{\eta}{G}$ \cite{frenkel}. This was later confirmed by experiments.

Eq. (\ref{a1}) can be written as

\begin{equation}
\frac{dv}{dy}=\frac{ds}{dt}=\frac{1}{\eta}AP
\label{a3}
\end{equation}

\noindent where $A$ is the operator

\begin{equation}
A=1+\tau\frac{d}{dt}
\label{a2}
\end{equation}

(\ref{a1})-(\ref{a3}) enable us to generalize $G$ to allow for long-time hydrodynamic flow \cite{frenkel}: noting that if $A^{-1}$ is the reciprocal operator to $A$, (\ref{a3}) can be written as $P=\eta A^{-1}\frac{ds}{dt}$. Because $\frac{d}{dt}=\frac{A-1}{\tau}$ from (\ref{a2}), $P=G(1-A^{-1})s$. Comparing this with the solid-like equation $P=Gs$, we see that the presence of hydrodynamic viscous flow is equivalent to the substitution of $G$ by the operator

\begin{equation}
G\rightarrow G(1-A^{-1})
\label{operator}
\end{equation}

Similarly, $\eta$ can be generalised to account for short-term elasticity: Eqs. \eqref{a3}-\eqref{a2} are equivalent to

\begin{equation}
\frac{1}{\eta}\rightarrow\frac{1}{\eta}\left(1+\tau\frac{d}{dt}\right)
\label{sub2}
\end{equation}

Eq. \eqref{sub2} can be used to generalise the Navier-Stokes equation for shear velocity field and obtain an equation governing the propagation of shear waves in viscoelastic liquids which gives gapped momentum states \cite{pre,physrep}. Eq. \eqref{operator} can be used to generalise the elastic stress-strain constitutive equation and obtain the same \cite{pre}.

Let us now consider a response to a periodic perturbation such as an external force $P\propto e^{i\omega t}$ as before. Then, Eq. \eqref{operator} gives $G(\omega)=G\frac{1}{1+\frac{1}{i\omega\tau}}$ \cite{frenkel}. We now write ${\rm Re}~G(\omega)=G\frac{\omega^2\tau^2}{1+\omega^2\tau^2}$, or

\begin{equation}
G(\omega)=G(1-F)
\label{etag1}
\end{equation}

\noindent where

\begin{equation}
F=\frac{1}{1+\omega^2\tau^2}
\label{factor}
\end{equation}

\noindent and where we dropped Re for simplicity.

Eqs. \eqref{etag1}-\eqref{factor} describe liquid viscoelasticity and its two limiting regimes $\omega\tau\ll 1$ in Eq. \eqref{cond} and $\omega\tau\gg 1$ in a transparent way. The hydrodynamic regime $\omega\tau\ll 1$ gives $G(\omega)=0$ as expected. The opposite solid-like elastic regime $\omega\tau\gg 1$ gives the expected Re $G(\omega)=G$.

The viscoelastic function $F$ in Eq. \eqref{factor} changes from $F=1$ in the hydrodynamic regime $\omega\tau\ll 1$ where the energy is entirely viscous and particle physically exchange places to $F=0$ in the opposite solidlike regime $\omega\tau\gg 1$ where all energy is entirely elastic and no exchanges operate. In time domain, this behavior is described by the function $F=1-e^{-t/\tau}$ (the exponential form follows from $P\propto e^{-t/\tau}$ in Eq. \eqref{a1} when $s=0$): $F=0$ at $t\ll\tau$ crosses over to $F=1$ at $t\gg\tau$. $F$ can therefore be used to account for the dynamical effects of particles indistinguishability such as the exchange energy. For example, writing the exchange energy $J$ as

\begin{equation}
J=J_0F
\label{newex}
\end{equation}

\noindent where $J_0$ is given in Eq. \eqref{exchange}, gives (a) $J=J_0$ in the regime $\omega\tau\ll 1$ ($t\gg\tau$) where particle exchange and exchange energy are operative and (b) $J=0$ in the regime $\omega\tau\gg 1$ ($t\ll\tau$) where particle exchange is inactive.

The above discussion applies to gases, with the proviso that the mechanism of particle exchanges becomes different in gases where particles fly in straight lines before undergoing collisions. In order for a pair of gas particles to exchange places, a particle needs to change its course and end up where the other particle was before. Hence, particle exchanges are inoperative during time $t$ shorter than the collision time $\tau_c$. In gases, $\tau_c=\frac{l}{v}$, where $l$ is the particle mean free path and $v$ is thermal velocity \cite{frenkel}. Recent work \cite{karen} quantified the transient regime in the gas where the product state of two particles far away from each other transitions to the symmetrised state due to wavefunction overlap.

The dynamical viscoelastic picture of quantum indistinguishability can be put in a wider context of understanding fluids. Traditionally, fluids were discussed in the hydrodynamic regime $\omega\tau\ll 1$ \cite{landau1}. Frenkel's theory \cite{frenkel} opened up a way to discuss important liquid properties in the opposite solid-like high-frequency regime $\omega\tau\gg 1$. A fairly recent realisation is that this regime is the key to understanding basic liquid dynamical and thermodynamic properties such as transverse phonons, energy and heat capacity \cite{ropp}. Here, we have begun discussing the implications of this high-frequency regime $\omega\tau\gg 1$ for quantum indistinguishability and ensuing effects.

From the point of view of fundamental understanding, it is interesting to discuss the experimental implications of dynamical quantum indistinguishability. Lets assume that the wave function starts evolving in response to changing external parameters such as temperature or pressure at time $t=t_0$ and that an observable is measured in the time window between $t_0$ and $t=t_0+t$, where $t\ll\tau$ so that the system is out of equilibrium. In this state, the wave function is a product rather than a symmetrised product \cite{l2}. The above discussion implies that an experiment would not show effects of quantum indistinguishability and statistics in this non-equilibrium state, including the formation of a condensate \cite{bec1,bec2} or Fermi surface \cite{zoran} in cold gases or a condensate in the quantum fluid \cite{condens}, as it wouldn't in their respective high-pressure crystal phases where particle exchanges are inoperative. Instead, BEC, Fermi surface and other statistics-related effects are predicted to be seen at longer time only (this time may considerably exceed $\tau$ to enable a macroscopically large number of exchange effects corresponding to the symmetrisation of the wave function, because $\tau$ marks the onset of particle exchanges only). We note that this long time corresponds to an equilibrium state where equilibrium Bose or Fermi distributions are considered to explain the above effects.

In the equilibrium state, the dynamics of particle exchanges and $\tau$ are related to the degree of overlap of particle wave functions and associated energy terms such as the exchange energy $J_0$ in Eq. \eqref{exchange} \cite{az}. For example, the transition rate $\Gamma$ in a system of two electrons between the state $|i\rangle=\phi_1(r_1)\phi_2(r_2)$ and the state with particles swapped, $|f\rangle=\phi_1(r_2)\phi_2(r_1)$, is $\Gamma\propto|\langle f|U|i\rangle|^2$ according to the Fermi golden rule, where $U$ is the interaction operator and $\langle f|U|i\rangle$ is its matrix element $\langle f|U|i\rangle=\int U\phi_1(r_1)\phi_1^*(r_2)\phi_2(r_2)\phi_2^*(r_1)dV_1dV_2$. This is the exchange energy $J_0$ setting the energy levels of two electrons and depending on the overlap of the wave functions $\phi_1(r_1)$ and $\phi_2(r_2)$ in Eq. \eqref{exchange}. In this simple model, $\Gamma\propto J_0^2$. $J_0$ depends on $U$, implying that $\tau$ and $J_0$ are related as well because (a) $\tau$ depends on the activation energy barrier $\tilde{U}$ and (b) $U$ and $\tilde{U}$ are closely related. Unless tunneling is involved, $\tau$ can additionally depend on other parameters such as temperature.

A related important effect from the fundamental point of view is the {\it transition} between regimes $t\ll\tau$ and $t\gg\tau$ in quantum systems. The importance of observing such a transition is supported by the interest generated by the experimental observation of Bose-Einstein condensation in cold gases \cite{bec1,bec2} and other quantum-statistical effects in boson and fermion systems \cite{l2}. In the proposed transition, one can observe the dynamical emergence of quantum indistinguishability, quantum statistics and its consequences such as BEC, Fermi surface and other ensuing properties. The experiments may be easier to perform in cold gases where relaxation times are relatively long and where the transition can be systematically studied in a given system at different $\tau$ (time between collisions). We have recently discussed specific details of such an experiment \cite{az}. This can guide new experiments in non-equilibrium quantum systems that have been of interest recently.

In summary, we took a first step to discuss the dynamical effect of quantum indistinguishability in fluids, formulated a quantitative criterion for observing its implications and proposed a viscoelastic function describing long-time and short-time regimes where the implications of indistinguishability are operative and inoperative. We proposed an experiment to observe a transition to the state where the implications of indistinguishability become inoperative.


\end{document}